\documentclass[aps,twocolumn,prb]{revtex4}
\usepackage{graphicx}
\usepackage{epsfig}
\usepackage{amsmath}
\usepackage{graphics}
\begin{document}

\title{First principle calculations of conductance within plane wave basis set
via nonorthogonal Wannier-type atomic orbitals}

\author{Zhenyu Li}
\author{D. S. Kosov} 

\affiliation{Department of Chemistry and Biochemistry,
 University of Maryland, College Park, 20742, USA}


\begin{abstract}
We present a plane wave/pseudopotential implementation of the 
method to calculate 
electron transport properties of nanostructures. The conductance 
is calculated via the Landauer formula within  formalism of Green's 
functions.
Nonorthogonal Wannier-type atomic orbitals are obtained by the sequential 
unitary rotations of  virtual and occupied Kohn-Sham orbitals,
which is followed by two-step variational localization. 
We use these non-orthogonal Wannier type atomic orbitals to partition the 
Kohn-Sham Hamiltonian into electrode-contact-electrode submatrices.
The electrode parts of the system are modeled by two  metal clusters 
with additional Lorentzian broadening of discrete energy levels. 
We examined our implementation by modeling the transport properties of Na 
atomic wires. Our results indicate that with the appropriate level broadening 
the small cluster model for contacts reproduces odd-even oscillations of the 
conductance as a function of the nanowire length.
\end{abstract}


\maketitle

\section{INTRODUCTION}
The last decade has witnessed a remarkable miniaturization of conventional microelectronic devices.
 If this trend is to continue, elements of microelectronic circuits, 
e.g. transistors and contacts, will soon shrink to the size of  a single molecule. 
One of the major goals in nanotechnology is the construction, measurement and modeling of electronic circuits 
in which molecular systems act as conducting elements \cite{nitzan0384}. 
Accurate and reproducible measurements of current-voltage characteristics 
have been recently reported for atomic wires and
 single molecules \cite{reed9752, cui0171, xu0321}.  The experimental progress has been accompanied by 
considerable advances towards density functional theory based 
calculations of transport properties of nanostructures. This activity has been largely spurred on
by  development of several electronic structure codes for the first principle transport calculations.
\cite{taylor0107,brandbyge0201,xue0251,evers04}

The prerequisite for non-equilibrium Green's function calculation of conductance is
the partitioning of the Kohn-Sham (KS) Hamiltonian into left/right electrodes 
and contact regions. Such partitioning is straightforward if 
one expands the KS wavefunctions as liner combinations of atomic 
orbitals 
\cite{taylor0107,brandbyge0201, xue0251,evers04} but it becomes a formidable 
theoretical problem
if the plane waves are used for a representation of the KS orbitals.
 One of the aims of
this paper is development of a theoretical scheme to the partition of the KS
Hamiltonian within plane wave basis set.

It has been recently proposed 
 that Wannier functions can be used to link  plane wave electronic structure  
and Green's function transport calculations.\cite{calzolari0408, thygesen05XX} 
Wannier functions are localized in coordinate space and  are  obtained by 
an unitary transformation of the KS orbitals. 
The number of 
Wannier functions is typically much smaller than the dimension of 
Gaussian-type basis sets. The additional advantage is that, 
we  treat the electrode and the wire at the same level of theory 
within the Wannier functions basis whereas it is not possible if one uses 
Gaussian basis set. The quality of Gaussian-type atomic basis set can not 
be easily controlled, 
which makes the results of calculation sensitive to
 a particular choice of basis functions. 
 In principle, these problems do not exist for plane wave basis set, 
where its quality is simply controlled by a single parameter, the cutoff energy.\cite{martin}
By increasing the cutoff energy, we can always get converged results. 
There are two disadvantages which makes standard  Wannier functions 
inapplicable for the calculations of
transport properties. First, Wannier functions are defined for occupied KS orbitals
and if the same localization scheme is directly used for the virtual 
orbitals it typically leaves them
as delocalized 
as they were before the unitary transformation. Second,
the centers of these localized Wannier functions are not controllable before minimization. 
To address both issues, we have developed the localization technique which yields 
non-orthogonal Wannier-type atomic orbitals (NOWAOs) from the plane-wave based KS orbitals.  
NOWAOs are the maximally localized functions 
defined via the set of unitary transformations of occupied and virtual KS orbitals. Our scheme is
based upon the combination of two localization techniques: Thygesen-Hansen-Jacobsen partially occupied 
Wannier functions \cite{thygesen0505} and Mortensen-Parrinello non-orthogonal
localization.\cite{mortensen0131} Combined together, these two techniques are used to include the virtual KS 
orbitals and to shift the Wannier centers from bonds to atoms. 

One additional ingredient, which is  necessary for calculations of conductance, is the Green's 
function of the electrodes.  In this paper, we represent the leads by two small clusters with additional Lorentzian
broadening of the energy levels.

We have implemented the working equations within a plane-wave/pseudopotential code\cite{cpmd}
and we will demonstrate the numerical accuracy of implementation for some prototypical test examples.
 The remainder of the paper is organized as follows. In Section II, we describe the details of 
our method to calculate the conductance of nanostructures, 
and it application to 
sodium atomic wires is given in Section III. In Section IV, we conclude the paper.

\section{METHOD}
\subsection{Partitioning of the Hamiltonian}

\begin{figure}
\centerline{
\epsfig{figure=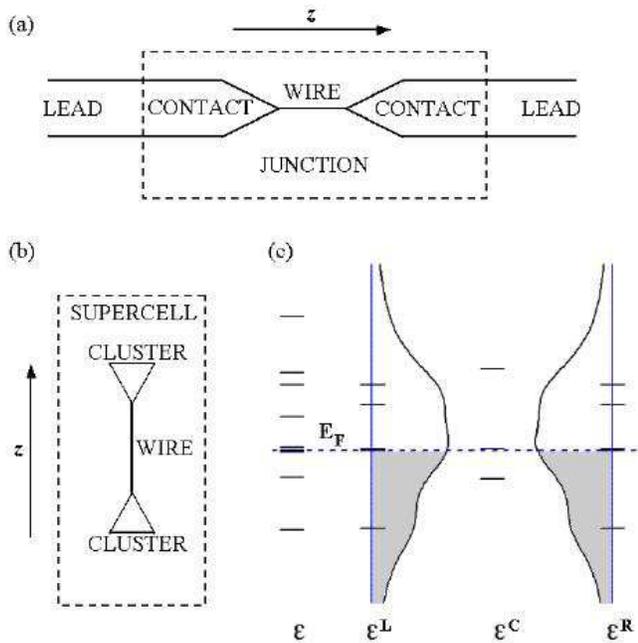,width=\columnwidth,angle=-0}}
\caption{ (a) Schematic representation of an atomic/molecular scale junction system for electronic transportation.  
(b) Simplified transportation model used in our method. (c) Energy levels for junction system containing
 three-atom sodium wire. Density of states 
obtained by Lorentzian broadening of the electrode levels are also shown. See text for details.} 
\label{fig:model}
\end{figure}

 We begin with the Kohn-Sham  equation for 
the entire nanowire junction
\begin{equation}
H |\psi_i\rangle =E_i|\psi_i\rangle\;,
\end{equation}
where $H$ is the Kohn-Sham Hamiltonian and $|\psi_i\rangle $ is the Kohn-Sham orbital.
The next step is the partitioning of the system into three parts: 
two electrodes and wire (typically atomic or molecular wire plus parts of the lead) as it is shown in Fig. \ref{fig:model}.
The partitioning of the system results into the partitioning of the Hamiltonian 
to seven  sub-matrices (left lead, contact, right lead and the lead-wire interactions) 
and it is performed by transforming the representation of the KS Hamiltonian from KS orbitals to atomically localized
basis sets:
\begin{equation}
|\psi_i\rangle = \sum_{n=1} U_{in} |w_n\rangle \;,
\end{equation}
where $U_{in} $ is a unitary transformation and  $ |w_n\rangle$  form atomically localized
complete basis set.
The unitary transformation $U_{in}$ is applied to the KS Hamiltonian 
\begin{eqnarray}
H&=&\sum_i |\psi_i \rangle E_i \langle \psi_i|=\sum_{nm} |w_n\rangle H_{nm} 
\langle w_m |
\label{h1}
\end{eqnarray}
with
\begin{equation}
H_{nm} =\sum_i U^*_{ni} E_i U_{im}
\label{h2} 
\end{equation}
 and yields the partitioning of the Hamiltonian into electrode-wire-electrode sub-matrices:
\begin{equation}
\mathbf{H}=\left(\begin{array}{ccc}
\mathbf{H}_L & \mathbf{H}_{WL}^{\dagger} & 0 \\
\mathbf{H}_{WL} & \mathbf{H}_W & \mathbf{H}_{WR} \\
0 & \mathbf{H}_{WR}^{\dagger} & \mathbf{H}_R \end{array} \right) \;.
\label{hamiltonian}
\end{equation}
If the basis set $|\omega_n\rangle $ is not orthogonal, the analogous partitioning should be
performed for
the overlap matrix $\mathbf{S}$  ($S_{mn} =\langle \omega_m | \omega_n \rangle$).
Matrices $\mathbf{H}$  and $\mathbf{S}$ can be definited if 
indexes $n$ and and $m$ in $H_{nm}$ (\ref{h1}) are associated with the atomic positions.
But this is not the case when periodic boundary conditions are employed and the KS
orbitals are expanded
 in plane waves:
\begin{equation}
\psi_i(\vec{r}) =\frac{1}{\sqrt{\Omega_{cell}}}\sum_{ \vec{G}} C_{i \vec{G}} \exp(i\vec{G} \vec{r}) \;,
\end{equation}
where  $C_{i\vec{G}}$ are the expansion coefficients. 
The plane waves $\exp(i\vec{G} \vec{r})$ do not have 
any reference to the atomic positions and therefore the
partitioning of the KS Hamiltonian into electrode-wire-electrode subspaces can not be performed directly
within the plane wave representation. Several groups have attempted to overcome this difficulty
by Wannier function representations of the  KS orbitals.\cite{calzolari0408,thygesen05XX}

\subsection{Non-orthogonal Wannier-type Atomic Orbitals}

Wannier functions are localized functions which span the same space as the eigenstates of a band or a group of bands.
Traditional Wannier functions are obtained by transforming Bloch representation to 
real space representation,\cite{wannier3791} in which Bloch vector $\vec{k}$ is substituted by the lattice vector $\vec{R}$ 
of the unit cell where the orbital is localized.
\begin{eqnarray}
|\vec{R}n\rangle&=&\frac{\Omega_{cell}}{(2\pi)^3}\int_{BZ}
{|\psi_{n\vec{k}}\rangle e^{i\phi_n(\vec{k})-i\vec{k}\cdot{\vec{R}}}d\vec{k}} \nonumber \\
&=&\frac{\Omega_{cell}}{(2\pi)^3}\int_{BZ}{\sum_{m}U_{mn}^{(\vec{k})}|\psi_{m\vec{k}}\rangle e^{-i\vec{k}\cdot{\vec{R}}}d\vec{k}}
\label{wan1}
\end{eqnarray}
where $U_{mn}^{(\vec{k})}$ is an arbitrary unitary matrix. The integration in eq.(\ref{wan1})
is done in reciprocal space within the whole Brillouin zone. The vector $\vec{k}$
equals zero for disordered systems 
like nanostructures or molecular wire junctions.
 In this case, the Wannier functions is defined via the
 unitary transformation of the KS orbitals\cite{berghold0040}
\begin{equation}
|\omega_n\rangle=\sum_m{U_{mn}|\psi_m\rangle}.
\end{equation}
There are several different schemes to define unitary matrix $U_{mn}$ and the choice the unitary 
transformation can be tailored to particular applications. Finding the maximally localized 
Wannier functions is pivotal for the partitioning of the Hamiltonian. 
Although there are several possible ways to define 
maximally localized Wannier functions, the method of 
minimization of the mean square spread stands out.\cite{marzari9747,souza0209}

Wannier functions are traditionally constructed only from the occupied 
KS orbitals. Occupied Wannier functions are located 
on the chemical bonds, which sometimes makes the partition 
of junction systems difficult. The additional complication is that the sum in eq.(\ref{h1}) 
runs over all KS orbitals (occupied and virtual) and due to the completeness requirement
it is necessary to include as many virtual KS orbitals in the consideration as possible.
Therefore, to be used in transport calculations Wannier functions 
should be constructed in such a way that  (a) they are 
 atomically localized and (b) they include 
both occupied and virtual KS orbitals.
The above two requirements are interconnected since to get atomic centered 
Wannier functions, we must anyway combine unoccupied anti-bonding states 
with occupied bonding states. 
The extraction of anti-bonding states from the entire set of 
virtual orbitals is difficult computational problem, since
unoccupied anti-bonding states are mixed with 
some scattering states originating from periodic 
boundary conditions.

Following Thygensen, Hansen and Jacobsen \cite{thygesen0505}
we begin the localization of the KS eigenstates by constructing
the linear combination of the virtual orbitals
\begin{equation}
\label{eq:anti}
|\phi_l\rangle=\sum_{m=1}^{N-M}{c_{ml}|\psi_{M+m}\rangle} \;,
\end{equation}
where $N$ is the number of KS orbitals and $M$ is the number 
of occupied states.
Partially occupied Wannier functions are written as\cite{thygesen0505}
\begin{equation}
\label{eq:U}
|\tilde{\phi}_n\rangle=\sum_{m=1}^M{U_{mn}|\psi_m\rangle}+\sum_{l=1}^L{U_{M+l,n}|\phi_l\rangle} \;,
\end{equation}
with $L$ being the number of unoccupied anti-bonding states. The optimal value of $L$ is yet 
to be determined.

We minimize the following localization functional
\begin{equation}
\Omega=\sum_n{[\langle\tilde{\phi}_n|r^2|\tilde{\phi}_n\rangle-\langle\tilde{\phi}_n|\vec{r}|\tilde{\phi}_n\rangle^2]}
\end{equation}
to choose a suitable unitary transformation $U_{mn}$ and to obtain the maximally-localized Wannier functions.\cite{marzari9747,souza0209}
Minimization of  functional $\Omega$ is equivalent to maximization of the functional
\begin{equation}
\Xi=\sum_n\sum_Iw_I|Z_{nn}^I|^2 \; ,
\label{xi}
\end{equation}
where matrix $\mathbf{Z}^I$ is defined as
\begin{equation}
Z_{mn}^I=\langle\tilde{\phi}_m|e^{-i\vec{G}_I\cdot\vec{r}}|\tilde{\phi}_n\rangle \;,
\end{equation}
with $\vec{G}_I$ and $w_I$  being the reciprocal lattice vectors and corresponding weights.\cite{berghold0040}  
For simple orthorhombic supercell, $I$ ranges from 1 to 3, corresponding to $x$, $y$, and $z$.
 In practical implementation,  $\mathbf{Z}^I$ is calculated by 
$\mathbf{Z}^I=\mathbf{U}^\dagger\mathbf{Z}_0^I\mathbf{U}$, with $\mathbf{Z}_0^I$ defined as
\begin{equation}
(Z_0^I)_{mn}=\langle\psi_m|e^{-i\vec{G}_I\cdot\vec{r}}|\psi_n\rangle \;.
\end{equation}

\begin{figure}
\centerline{
\epsfig{figure=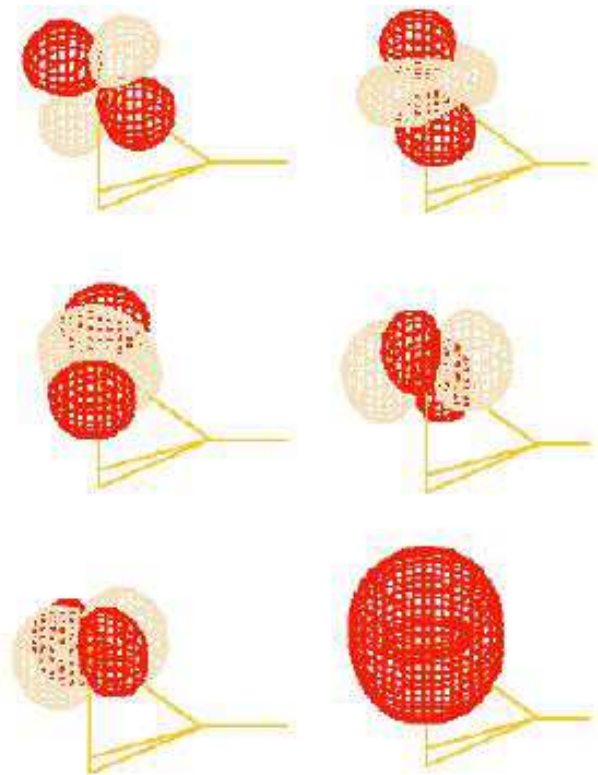,width=\columnwidth,angle=-0}}
\caption{ (Color online) Non-orthogonal Wannier-type atomic basis functions (NOWAOs) for one gold
 atom in a simple gold junction system. Only half of the junction is shown. }
\label{fig:nowbf}
\end{figure}

Analytical gradients of  functional $\Xi$ (\ref{xi}) are  necessary to perform effective maximization.
 If we write the unitary matrix at iteration $i$ 
as $\mathbf{U}_i=\mathbf{U}_{i-1}\exp{(-\mathbf{A})}$, 
then the gradients of functional $\Xi$ with respect to $\mathbf{A}$ can be approximated as
\begin{equation}
\left(d\Xi/dA\right)_{ij}=\sum_Iw_I[Z_{ji}(Z_{jj}^*-Z_{ii}^*)-Z_{ij}^*(Z_{ii}-Z_{jj})] \; .
\end{equation}
The gradient with respect to the coefficient matrix $c_{lm}$ is computed by the following formula\cite{thygesen0505}
\begin{eqnarray}
&&\left(d\Xi/dc^*\right)_{ij}= \\
&&\sum_Iw_I[[\mathbf{Z_0}\tilde{\mathbf{U}}\mbox{diag}(\mathbf{Z}^{\dagger})
+\mathbf{Z_0}^\dagger\tilde{\mathbf{U}}\mbox{diag}(\mathbf{Z})]\mathbf{U}^\dagger]_{N+i,N+j} \;,
\nonumber
\end{eqnarray}
where $diag(\mathbf{Z})$ is the diagonal part of matrix $\mathbf{Z}$, and $\tilde{\mathbf{U}}$ 
is the rotation matrix from KS orbitals to partly occupied Wannier functions with dimension $N\times(M+L)$.
The orthonormality constraint on matrix $\mathbf{c}$ is invoked through a set of Lagrange multipliers. 
The steepest descent method is used to maximize $\Xi$. After the maximization, the
 unoccupied anti-bonding states are obtained from the coefficient matrix $\mathbf{c}$ through Eq.(\ref{eq:anti}).

A final set NOWAOs is computed  via the additional rotation of
partially occupied Wannier functions (\ref{eq:U}):
\begin{equation}
\label{eq:V}
|\omega_n\rangle=\sum_{m=1}^{M+L}{V_{mn}|\tilde{\phi}_m\rangle}\;.
\end{equation}
Rotation matrix $V_{nm}$ is defined 
by minimizing the following function\cite{mortensen0131} independently for each $n$:
\begin{equation}
\Omega_n=\langle\omega_n|p(\vec{r}-\vec{R}_n)|\omega_n\rangle
\end{equation}
The weight function $p(\vec{r})$ is chosen in such a way that it has a minimum 
at $r=0$ to localize each $|\omega_n>$ around $\vec{R}_n$. 
Following Mortensen and  Parrinello\cite{mortensen0131} we select function $p(\vec{r})$ as
\begin{equation}
p(\vec{r})=\sum_{\alpha=x,y,z}\left[{1-\cos(\frac{2\pi}{L}r_\alpha)}\right]\;.
\end{equation}
$\Omega_n$ can also be written in the matrix form:
\begin{equation}
\Omega_n=(\mathbf{V}^\dagger\mathbf{P}^{(n)}\mathbf{V})_{nn}=\vec{v}_n^\dagger\mathbf{P}^{(n)}\vec{v}_n \;,
\end{equation}
where $\vec{v}_n$ is the $n$th column of $\mathbf{V}$, and
\begin{equation}
P^{(n)}_{ij}=\langle\tilde{\phi}_i|p(\vec{r}-\vec{R}_n)|\tilde{\phi}_j\rangle
\end{equation}
with $|\tilde{\phi}>$ being the set of partially occupied Wannier functions obtained by 
maximization of $\Xi$. Matrix element $\mathbf{P}^{(n)}$ are calculated only once 
and stored for every$n$. The minimum of $\Omega_n$ is obtained when $\vec{u}_n$ is 
equal to the normalized eigenvector corresponding to the smallest eigenvalue of 
$\mathbf{P}^{(n)}$. If we need several NOWAOs for a single atomic 
site, the corresponding number of smallest eigenvectors should be chosen. 
The number of anti-bonding states necessary for the localization ($L$ in eq.(\ref{eq:U})
can now be computed by the following formula:
\begin{equation}
L=N_AN_{LE} -M \;,
\end{equation}
where $N_A$ is the number of atoms in the system and $N_LE$ is the 
number of the lowest eigenstates included into Mortensen-Parrinello localization.
For example, 
for gold atom, which is typical electrode material in molecular electronics,
 five $d$-type and one $s$-type NOWAOs per atom are need. In Fig. \ref{fig:nowbf}, 
we plot the generated six NOWAOs for a gold atom in a simple gold wire junction system. 
We can clearly see that these six NOWAOs reflect the $s$ and $d$ characters of gold atom.

\subsection{Conductance Formula}
The starting point for the conductance calculations is the Landauer formula\cite{datta97}
\begin{equation}
G=\frac{2e^2}{h}T(E_F),
\end{equation}
where $T$ is the transmission function and $E_F$ is the Fermi energy of the electrodes. 
Having obtained the partitioned Hamiltonian (\ref{hamiltonian}), 
we can compute transmission as the trace of Green's function $\mathbf{G}$ and coupling 
matrices $\mathbf{\Gamma}_{L/R}$:\cite{xue0251}
\begin{equation}
T(E)=\mbox{Tr}[\mathbf{\Gamma}_L(E)\mathbf{G}(E)\mathbf{\Gamma}_R(E)\mathbf{G}^{\dagger}(E)].
\end{equation}
The matrices $\mathbf{G}$ and $\mathbf{\Gamma}_{L/R}$ are expressed by the matrix blocks of 
the Hamiltonian $\mathbf{H}$, overlap $\mathbf{S}$, self energy  $\mathbf{\Sigma}$:
\begin{equation}
\mathbf{G}(E)=[E\mathbf{S}_W-\mathbf{H}_W-\mathbf{\Sigma}_L-\mathbf{\Sigma}_R]^{-1}
\end{equation}
and
\begin{equation}
\mathbf{\Gamma}_{L/R}(E)=i[\mathbf{\Sigma}_{L/R}-\mathbf{\Sigma}_{L/R}^{\dagger}].
\end{equation}
The self-energies $\mathbf{\Sigma}_{L/R}$ are defined via the Green's function 
of the left and right electrodes $\mathbf{g}$ and the electrode-wire interactions:
\begin{eqnarray}
&&\mathbf{\Sigma}_{L/R}(E)= \\
&&(E\mathbf{S}_{WL/R}-\mathbf{H}_{WL/R})\mathbf{g}_{L/R}(E)(E\mathbf{S}^{\dagger}_{WL/R}
-\mathbf{H}_{WL/R}^{\dagger}).
\nonumber
\end{eqnarray}
It is not possible to include the entire leads in the practical calculations.
The interaction between the wire and the infinitely large leads is accounted 
by the self-energy.
Different theoretical models have been proposed 
to obtain $\mathbf{g}$. The most accurate schemes relay on the surface Green's function method. 
Even with some special techniques, such as transfer matrix\cite{lee8188,lee8197} and 
decimation technique\cite{sancho8551}, the computation of surface Green's function is  complex and time-consuming. 
Since  detailed experimental geometrical structures of leads and a nanowire are unclear in most cases,
it enables us to use the simpler models for the leads.
In this paper, the electrode is represented by a metal cluster with additional  energy level broadening\cite{tada0450}.
 This model has proven to
be very successful in estimation of the bulk density of states from small cluster calculations.  
\cite{lee8424,zhao0304}
 As shown in APPENDIX A, the Lorentzian broadening $\sigma$ 
in the electrode density of state is the same as is positive infinitesimal $\sigma$
in the electrode Green's function:
\begin{equation}
\mathbf{g}= [(E+i\sigma)\mathbf{S}_{L/R}-\mathbf{H}_{L/R}]^{-1}\;.
\end{equation}

\section{Test results}
\subsection{Computational methods}
To illustrate the performance of our method we computed transport properties of Na nanowire.
Nanowires of metal atoms have recently attracted much attention because of their 
fundamental and technological importance. In particular, sodium atomic wire has been
 studied both experimentally\cite{krans9567, krans9628, yanson9944} and  
theoretically.\cite{lang9757, sim0103, tsukamoto0202, lee0409, khomyakov0402}  
It was  found that the conductance of Na wires exhibits even-odd oscillation  as a function of the
number of atoms in the wire. The conductance for odd sodium atom numbered wire is close 
to the quantum of conductance $G_0$ ($2e^2/h$), while the conductance for even sodium 
atom numbered wires is smaller than $G_0$.
Different implementation and junction models lead to different the values of the conductance for
even numbered nanowires (0.5 -- 0.9 $G_0$).\cite{lang9757, sim0103, tsukamoto0202, lee0409, khomyakov0402} We use Na atomic nanowire as a proving ground for our implementation
and we aim to reproduce the odd-even oscillation and values of the conductance.

The calculations were performed using our implementation of
the formalism presented here in the
 CPMD package \cite{cpmd}. All  systems
were treated employing periodic boundary conditions and the KS orbitals
were expanded in plane waves (50 Ry cutoff) at the $\Gamma$ point of the Brillouin zone.
We used local density approximation for the exchange and correlation functional and Stumpf, Gonze, and Schettler
pseudopotentials\cite{gonze9103} for core electrons.
The system is simulated by a cluster in a large supercell. The size of  supercell 
is chosen in such a way that the distance between the nearest atoms in the neighboring cells is larger 
than 8.5\AA, so that the interaction between supercell images is negligible. 
An extensive set of the KS virtual orbitals is computed via Lanczos diagonalization\cite{pollard9342}
 to ensure that all possible unoccupied anti-bonding states are included. 
 To speed up convergence of the self-consistent iterations, free energy functional\cite{alavi9499} 
is used with the electronic temperature $T=300$ K. Since sodium
is a single valent atom, only one NOWAO per Na atom is constructed from the KS orbitals.

\begin{figure}
\centerline{
\epsfig{figure=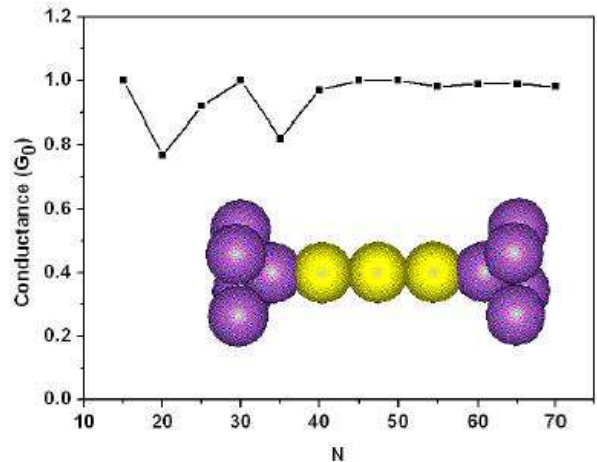,width=\columnwidth,angle=-0}}
\caption{Conductance of three-atom Na atomic wire as a function of number of Kohn-Sham orbitals calculated to construct the NOWAOs. The inset shows the geometry of this junction system.} 
\label{fig:nstate}
\end{figure}

The whole system is divided into three parts: left electrode, central wire, and right electrode.
 The electrode part is obtained by cutting a few atoms from Na (001) surface. In 
particular, as shown in the inset of Fig. \ref{fig:nstate}, we cut  a five-atom 
cluster, which is composed of  square four-atom base and apex atom. The geometry
of this five-atom cluster is fixed to the bulk values. The wire part is a 
single chain of Na atoms, where the distance between the atoms is constrained to 
the nearest neighbor distance in the bulk system. The distance between the electrode 
part and the wire part is optimized. The optimized value $d$ is listed in 
Table \ref{tbl:conductance} and it shows small $\sim 0.1-0.2$ \AA\ odd-even oscillation
as the length of the wire varies. We use optimized value of electrode-wire separation in 
all our  calculations.

Let us take a three-atom Na wire as an example to illustrate  
conductance calculations by the method we described in the previous section. 

First,  
electronic structure of the whole system is calculated, and a set of KS 
orbitals is obtained. The number of KS orbitals should be large enough 
to include  all unoccupied anti-bonding states. 
Fig.\ref{fig:nstate} shows the conductance as a function of the number 
of KS orbitals included in the localization procedure to define 
NOWAOs. The deviation of the conductance from the correct value (1 $G_0$) vanishes rapidly
as the number of virtual KS orbitals  is increased. Our test calculations illustrate
 the number of the virtual KS orbitals can be adjusted to achieve any desired level of accuracy in
the conductance calculations. The results on Fig.\ref{fig:nstate} demonstrate that 45 virtual KS orbitals
are sufficient to get the converged result for three atom Na wire. 

Second, this set of KS
orbitals is used to construct NOWAOs, with which Hamiltonian matrix $\mathbf{H}$ 
and overlap matrix $\mathbf{S}$ are calculated. By solving the corresponding generalized 
eigenvalue problem, we get a set of energy levels $\mathbf{\epsilon}$ for the whole 
junction system, as shown in the left column of Fig. \ref{fig:model}c. 
The differences between these eigenlevels and the Kohn-Sham energies are 
very small, especially for occupied state. At the same time, partition of $\mathbf{H}$ and $\mathbf{S}$ 
is implemented, and the energy levels of electrodes, $\epsilon^L$ and $\epsilon^R$, are obtained
by solving the generalized eigenvalue problem for the corresponding submatrices.

%
Third, we introduce Lorentzian broadening of the left and right energy 
levels to continuous density of states, 
to compute the self-energy matrices. The 
broadening parameter $\sigma$ is chosen to be 0.025 Hartree here, and the effect 
of broadening parameter on conductance will be discussed later. The transmission probability is
computed by using section II.B equations.
The peak positions of the transmission curve
 can be considered as renormalized eigenlevels of the 
central three-atom Na wire coupled to the electrodes ($\epsilon^C$ in Fig. \ref{fig:model}).

In units of $G_0$, the conductance equals to the 
value of the transmission at  Fermi energy of the leads. Fermi energy for the 
electrode is not known {\it apriori} and it has to be computed within the approach. 
There are two schemes to compute Fermi energy of the electrodes. The first one 
determines the Fermi energy 
at which the integrated density of states should be equal 
to the number of electrons in the electrode cluster. \cite{tada0450}
The charge population of the electrode cluster is calculated by 
$\rho_{L/R}=\sum_{i\in{L/R}}(\mathbf{S}^{1/2}\mathbf{P}\mathbf{S}^{1/2})_{ii}$,
where $\mathbf{P}$ is the density matrix on NOWAOs basis set,  which is
obtained from the
generalized eigenvalue problem of the whole junction cluster system. The other one is more 
simple, we can just use the Fermi energy of the whole junction system. When the electrode 
clusters are very large, both methods become equivalent. For the five-atom model 
used here, we find the later method is preferable and it yields 1 $G_0$ value of the conductance 
of the three-atom Na wire system.

\subsection{Oscillation of Conductance}
 
\begin{figure}
\centerline{
\epsfig{figure=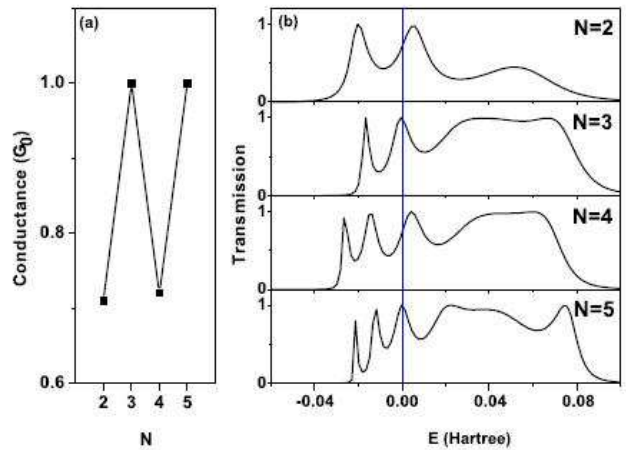,width=\columnwidth,angle=-0}}
\caption{(a) Conductance oscillation of sodium atomic wire with its length. (b) Transmission curves of the sodium atomic wires. Fermi energy is set to zero.}
\label{fig:trans}
\end{figure}

The most important feature of electronic current flow through Na atomic wires is 
that the conductance oscillates as a function of the number of atoms in the wire
We calculate the conductance of Na atomic wires  with length range from 2 to 5 atoms. The odd-even 
oscillation of the conductance is well 
reproduced as shown in Fig. \ref{fig:trans}a. The conductance for $N=3$ and $N=5$ is  one unit conductance 
$G_0$, while the conductance for $N=2$ and $N=4$ is about 0.7 $G_0$. The transmission curves for these wire systems 
are plotted in Fig. \ref{fig:trans}b. We can see that the number of sharp peaks in the transmission curves, i.e. the number of resonant states, is equal to the number of atoms in the wire. 
The low temperature transport properties are mainly determined by the states near the Fermi energy. 
There is a resonant state at the Fermi energy for odd numbered Na wires, 
but there is no such state for even numbered Na wires. 
This is exactly the reason of the even-odd oscillation behavior of the conductance.\cite{sim0103,lee0409}

\subsection{Eigenchannels}

Since there are multiple peaks in the transmission curve, 
it is interesting to see if sodium wires are single channel 
conductor as we expected for monovalent atomic chain. By defining 
\begin{equation}
\mathbf{t}=\mathbf{\Gamma}_L^{\frac{1}{2}}\mathbf{G}^R\mathbf{\Gamma}_R^{\frac{1}{2}}
\end{equation}
the transmission can be written as $T(E)=\mbox{Tr}(\mathbf{t}\mathbf{t}^\dagger)$. We can decompose 
the transmission to the combination of some eigenchannels diagonalizing
the  matrix $\mathbf{t}\mathbf{t}^\dagger$.\cite{brandbyge9756} 
For sodium wires, we get only one non-zero eigenvalues at any energy, which 
is the manifestation of a single channel conduction mechanism.

\subsection{Electrode Cluster Model}

\begin{table}
\caption{Distance between electrode and wire $d$ (in \AA), and conductance of the 
Na atomic wire junction systems. The length of Na wire $N$ ranges from 2 to 5 atoms. 
The electrode part of the junction system is modeled by one ( E1 model ) 
or five (E5 model) Na atoms. Conductance for four different broadening 
parameter 0.02, 0.025, 0.03, and 0.04 Hartree are listed.   See text for more details.} \label{tbl:conductance}
\begin{tabular}{ccccccccccccc}
\hline\hline
 && \multicolumn{5}{c}{E1 Model} && \multicolumn{5}{c}{E5 Model}  \\
 \cline{3-7}\cline{9-13}
$N$ && $d$ & 0.02& 0.025 & 0.03& 0.04 && $d$ & 0.02& 0.025 & 0.03& 0.04 \\
\hline
2  && 3.004 & 0.33 & 0.43 & 0.50 & 0.57 &&  3.078 & 0.64 & 0.71 & 0.75 & 0.77 \\
3  && 3.070 & 0.99 & 0.99 & 0.98 & 0.97 &&  3.270 & 0.99 & 1.00 & 1.00 & 0.99 \\
4  && 3.024 & 0.65 & 0.81 & 0.92 & 1.00 &&  3.097 & 0.71 & 0.72 & 0.72 & 0.69 \\
5  && 3.042 & 0.99 & 0.99 & 0.99 & 0.99 &&  3.221 & 1.00 & 1.00 & 0.98 & 0.96  \\
\hline\hline
\end{tabular}
\end{table}

In our method, the  electrode is approximated by a small cluster with proper level broadening. 
Therefore it is important to understand how the results of our calculations are affected by the 
cluster size and by the choice of the broadening.

The size of the electrode cluster should be big enough, so that charge neutrality is 
maintained for the junction system. As it was discussed in the sections III.A and  III.B 
the  five-atom electrode  model 
is sufficient to simulate the  transport behavior of sodium wires in 
our method.\cite{clustersize} It is already a very small size comparing to typically used
models for electrodes, 
but it is interesting to see if a smaller cluster  still works in our method. For this purpose, we also 
check the simplest one-atom model. The results are listed in Table \ref{tbl:conductance}. Comparing 
to the five-atom model, the oscillation strength of the optimized distance between electrode and wire is 
smaller. We can see that the conductance for one-atom model still exhibits the even-odd oscillation, 
but the conductance of even numbered wires are not very stable anymore.  This result indicates that 
the one-atom electrode model provides the oversimplified description of the system.

It is also important to test if the broadening parameter strongly effects results of the transport calculations. 
We analyzed the conductances for all wires by varying the broadening parameter $\sigma$ from
from 0.02 to 0.04 Hatree, and list the result in Table \ref{tbl:conductance}. We find that 
for large electrode model,  the conductance is not very sensitive to the variations of $\sigma$, especially
 for odd-numbered wire system. For one-atom electrode model, the conductance changes significantly
as $\sigma$ varies. This is because there is only one electrode energy level for one-atom 
model, and the shape of the density of states at $E_F$ strongly depends on the broadening parameter.

There is another conductance calculation method 
for single channel nanowire.\cite{datta9714,sim0103} 
It is based on Friedel sum rule and also relays 
on the eigenlevel broadening technique to represent continuum of states in the contact. In that method, the 
eigenvalues of the whole junction system $\epsilon$ are broadened to obtain density of states, thus the broadening parameter 
should  be smaller than the resonance spacing but larger than single particle level spacing for 
electrode. To satisfy this constraint, the spacing of $\epsilon^L$ and $\epsilon_R$ should be much 
smaller than that of $\epsilon^C$, and therefore much larger electrode clusters should be used
in Friedel sum rule based calculations. 
In our method, 
however, if the broadening parameter is comparable to the resonance spacing, we still get very 
accurate results. Therefore, much smaller metal clusters can be used to model electrodes
within our implementation.

\section{Conclusions}

We have developed and implemented a plane wave based method to calculate 
the conductance of nanostructures. The fundamental quantity in the
present implementation is NOWAOs which are obtained by the multi-step
localization of KS orbitals. NOWAOs are used to partition KS Hamiltonian 
to electrode-wire-electrode submatrices that is necessary step for
the Green's function based conductance calculations.
The electrode parts are modeled by  small clusters with proper 
broadening of their eigenlevels and this model is especially suitable 
for systems with very limited 
knowledge of lead-wire bonding structure. 
Transport properties of sodium  nanowires are 
studied by this method and the odd-even oscillations  
of the conductance are  reproduced.

\begin{acknowledgments}

The authors are grateful to F. Evers, M. Gelin and K. S. Thygesen for helpful discussion.

\end{acknowledgments}

\begin{appendix}
\section{Lorentzian broadening of electrodes levels }
 The density of states of the leads is related to imaginary part of Green's function by
\begin{equation}
\mbox{Im}\mathbf{g}(E)=-\pi\mathbf{D}(E)
\end{equation}
and  real part of Green's function can then be obtained by Kramers-Koning relation
\begin{equation}
\mbox{Re}\mathbf{g}(E)=\frac{1}{\pi}P\int_{-\infty}^{+\infty}{\frac{\mbox{Im}\mathbf{g}(\omega)}{\omega-E}d\omega} \;.
\end{equation}
Here $P$ stands for the Cauchy principal value integral. 
 For the widely used Gaussian broadening,\cite{tada0450} 
this procedure is numerically complicated (Romberg integration technique) and time-consuming. 
We use the Lorentzian broadening instead and we will show in this appendix that the
broadening parameter in Lorentzian for the density of states is the same as the infinitesimal 
parameter in the non-interacting electrode Green's function. 
The Lorentzian broadening function is written as
\begin{equation}
D(E)=\frac{1}{\pi}\sum_i{\frac{d_i\sigma}{(E-\epsilon_i)^2+\sigma^2}}\;,
\end{equation}
where $\epsilon_i$ are the eigenlevels for the left or right electrode cluster.
The Cauchy principal integral can be obtained analytically
for the Lorentzian density of states and it yields the following Green's function matrix for the leads
\begin{equation}
{g}_{ij}(E)=\frac{\delta_{ij}}{(E-\epsilon_i)^2+\sigma^2}(E-\epsilon_i-i\sigma) 
\end{equation}
or in matrix form
\begin{eqnarray}
\mathbf{g}(E)&=&\frac{\delta_{ij}}{(E-\mathbf{H})^2+\sigma^2}(E-\mathbf{H}-i\sigma) \nonumber \\
&=& 
[(E+i\sigma)\mathbf{I}-\mathbf{H}]^{-1}\;.
\end{eqnarray}

\end{appendix}


\begin{thebibliography}{99}
\bibitem{clustersize} We have also test the 10-atom cluster model, which is obtained by adding another 
layer to the electrode from the Na (001) surface. With smaller broadening parameters, the bigger model gives 
similar results aas the 5-atom cluster model discussed in the text.
\end{thebibliography}
\end{document}